\documentclass[preprint2]{aastex} 

\newcommand{\beq}{\begin{equation}}
\newcommand{\eeq}{\end{equation}}
\def\beqa{\begin{eqnarray}}
\def\eeqa{\end{eqnarray}}
\newcommand{\gam}{\langle\sqrt{1 - v^2}\rangle}
\newcommand{\xv}{{\bf x}}

\newcommand{\kv}{{\bf k}}

\begin{document}

\title{Interacting Dark Matter and Dark Energy}

\author{Glennys R. Farrar$^{1,2}$
and P. J. E. Peebles$^{1}$}

\affil{$^1$Joseph Henry Laboratories, 
Princeton University,
Princeton, NJ 08544}

\affil{$^2$ Center for Cosmology and Particle
Physics,
New York University, New York NY
10003}

\begin{abstract}
We discuss models for the cosmological dark sector in which the
energy density of a scalar field approximates Einstein's
cosmological constant and the scalar field value determines the
dark matter particle mass by a Yukawa coupling. A model with one dark matter family can
be adjusted so the observational constraints on the cosmological
parameters are close to but different from what is predicted by
the $\Lambda$CDM model. This may be a useful aid to judging how
tightly the cosmological parameters are constrained by the new
generation of cosmological tests that depend on the theory of 
structure formation. In a model with two families of dark matter
particles the scalar field may be locked to near zero mass for one
family. This can suppress the long-range scalar force in the dark
sector and eliminate evolution of the effective cosmological
constant and the mass of the nonrelativistic dark matter
particles, making the model close to $\Lambda$CDM, 
until the particle number density becomes low enough to
allow the scalar field to evolve. This is a useful example of the possibility for complexity in the dark sector. 
\end{abstract}


\section{Introduction}

The striking success of the $\Lambda$CDM model in fitting the
precision WMAP measurementsÊ of the anisotropy of the 3~K thermal
cosmic background radiation and the other cosmological tests (\citet{wmap} and references therein)
shows this cosmology is a useful approximation to the physics of
the dark matter and dark energy.Ê However, it is not difficult to imagine 
 more complicated physics in the dark sector. If the
physics differs from $\Lambda$CDM enough to matter, it will be
manifest as anomalies in the fits to the observations. It is
prudent to anticipate this possibility, by
exploring models for more complicated physics in the dark sector.

The starting idea for the physics under discussion in this paper is that the dark matter (DM) 
particle mass may be determined by its interaction with a scalar field whose energy density
is the dark energy (DE).\footnote{We do not consider whether such a scenario can emerge naturally in supersymmetry or axion models. Perhaps it requires a new view of dark matter.} We explore the physics and astrophysics of models for this extension of  $\Lambda$CDM under the simplifying assumptions of general relativity theory, standard physics in the visible sector, and, in the dark sector, a Yukawa coupling of the DE field to the DM particles. Much of the physics, as summarized in \S 3, is in the literature, but we have not seen it all collected and applied to the astrophysics. The example application in \S 4, which assumes a single DM family, allows parameter choices that make the model predictions viable but different from $\Lambda$CDM. The example in \S 5, with two DM families, allows an interesting  mixture of near equivalence to $\Lambda$CDM at early times and complicated departures from this model at late times.  

The line of ideas in this topic has a long  history, which informs assessments of where we are now. In \S 2 we present our selection of the  main steps in the historical development.  

For convenient reference we write down here the forms we will be considering for the action in the dark sector, as a sum of two terms.  The first is the familiar DE model, 
\beq
 S_{\rm DE} = \int d^4x\sqrt{-g}\left[ {1\over 2}\phi _{,\nu }\phi ^{,\nu
} - V(\phi )\right] ,\label{eq:se}
  \eeq 
where the function $V(\phi )$
of the classical real DE scalar field $\phi$ is chosen so the
field stress-energy tensor approximates the effect of the
cosmological constant $\Lambda$ in Einstein's field equation. In
numerical examples we use the power law potential, 
\beq 
V(\phi ) =
K/\phi ^\alpha ,\label{eq:V}
 \eeq 
 where $K$ is a positive constant and the constant $\alpha$ may be positive or negative. The DM term, in the form  used in much of our discussion, is 
\beq 
S_{\rm DMf} = \int
d^4x\sqrt{-g}\left[ i\bar\psi\gamma\cdot\partial\psi - y (\phi
-\phi _\ast)\bar\psi\psi\right] . \label{eq:sf}
 \eeq 
The subscript, DMf, indicates this is the action written in terms of
the wave function $\psi$ for a spin-1/2 DM field. In the Yukawa
interaction term, $y$ is a dimensionless constant and the constant
$\phi _\ast$ has units of energy (withÊ $\hbar = 1 = c$). If $\phi
_\ast$ in equation~(\ref{eq:sf}) is negligibly small, the entire
particle mass is due to its interaction with the field $\phi$.  This seems particularly attractive because we may need this field anyway, to account for the DE. The interaction between the DM and DE allows the DM particle mass to be variable, producing a long-range nongravitational interaction  in the dark sector. Both effects can be suppressed by the presence of a scond DM family with a different value of $\phi _\ast$, as we discuss in \S 5, or by suitable choices of $\phi _\ast$ and $V(\phi )$  for one family (\S 4). 

For completeness one might also consider the analogue of equation~(\ref{eq:sf}) for scalar DM particles. When $\phi _\ast$ is negligibly small the analogous form is\footnote{The constant $\phi _\ast$ in equations~(\ref{eq:sf}) and~(\ref{eq:sb}) is the sum of ``bare'' and  renormalization parts. For generality one would add a constant to $y^2(\phi -\phi _\ast)^2$ in equation~(\ref{eq:sb}). We ignore considerations of naturalness in the choice of these constants. Issues of naturalness and renormalization  plague other
aspects of cosmology and fundamental particle physics, as in the meaning of equation~(\ref{eq:V}) within quantum field theory, and most notably the value of the vacuum energy density.} 
 \beqa
S_{\rm DMb} = \int d^4x\sqrt{-g}
\Bigl[ {1\over 2}\chi _{,\nu }\chi ^{,\nu} - \nonumber \\
 \qquad {1\over 2} y^2(\phi -\phi _\ast)^2\chi ^2\Bigr] .\label{eq:sb}
 \eeqa

In the limiting situation that is thought to apply to cosmology, where 
the DM particle de Broglie wavelengths are much smaller than the characteristic length scale of variation of the DE field, both DM actions (eqs. [\ref{eq:sf}] and [\ref{eq:sb}]) are  equivalent to the model of a classical gas of point-like particles with action\footnote{It will be recalled that the exclusion principle affects initial
conditions -- the occupation numbers in single particle phase
space -- but not the equation of motion in the particle limit.}
\beq 
S_{\rm DMp}Ê =Ê - \sum _i \int y |\phi (x_i) - \phi
_\ast|ds_i. \label{eq:sp} 
\eeq 
The invariant interval along the
path $x^\mu _i(t)$ of the $i^{\rm th}$ particle is
$ds_i=\sqrt{g_{\mu\nu }dx_i^\mu dx_i^\nu}$. The DE particle mass
is $m_{\rm eff}=y |\phi -\phi _\ast|$. The absolute value in equation~(\ref{eq:sp}) has to be a prescription (along with $y>0$),
but as we discuss in \S3 it is not needed in
equations~(\ref{eq:sf}) or~(\ref{eq:sb}). Equation~(\ref{eq:sp}) is a convenient
form for analyses of structure formation. 

\section{Remarks on the History of Ideas}

The particle action in equation~(\ref{eq:sp}), with a variable effective mass, appears in
Nordstr\"om's (1912) scalar field model for gravity in Minkowski
spacetime,  in the form  $L_i \propto e^{-\phi /\phi _\ast}ds_i$ (in our
notation). This form reappears in  Misner, Thorne \& Wheeler (1973), 
and it still is favored by many, in part because the
exponential is suggested by superstring theory. The functional form does not much affect the force generated by the particle-field interaction, but it affects the cosmic evolution of particle masses and the force between them. In our preliminary examination of possible alternatives to the $\Lambda$CDM model we prefer the linear form in
equation~(\ref{eq:sp}), 
because it translates to the familiar and simple Yukawa
interaction in the field action model in equation~(\ref{eq:sf}).
This linear coupling appears also (explicitly or as a particular
case) in many recent discussions of the possible interaction of DM and DE (e.g.  Casas, Garcia-Bellido \&\ Quiros 1992; Anderson \&\ Carroll 1997; Bean 2001, and references therein). 

The particle action with variable mass appears in the scalar-tensor gravity theory considered by Jordan (1955, 1959) and Brans \&\ Dicke (1961), when expressed in units chosen so the action for gravity is  the Einstein form (Fierz 1956; Dicke 1965). The units in this theory may be rescaled to standard local physics with constant masses and a generalized action for gravity, which is the form Jordan and Brans \&\ Dicke used to implement Dirac's (1938) idea that the  strength of the gravitational interaction may be small because it is rolling toward zero. 

Superstring scenarios led to the thought that  particle mass ratios (and other dimensionless constants) may be variable, and in particular  gravity physics and local physics in the visible sector may be close to standard while particle masses in the dark sector are more significantly variable (Damour, Gibbons \&\ Gundlach 1990). This allows interesting departures from standard cosmology within the tight constraints on gravity physics from precision tests in the visible sector.
Damour et al. (1990) work with a scalar-tensor theory for gravity physics,  a route taken in many subsequent papers. For clarity in our exploratory discussion of the dark sector physics we adopt general relativity theory, standard physics in the visible sector, and a single scalar field that fixes particle masses in the dark sector. 

The scalar field model for the DE in equation~(\ref{eq:se}) was introduced by Wetterich (1988) and Peebles \&\ Ratra (1988) (as reviewed in Peebles \&\ Ratra 2003). This DE model allows one to imagine that the effective cosmological constant is evolving to its ``natural'' value, $\Lambda =0$, and is small now because the universe is old, a natural extension of the ideas of Dirac (1938), Jordan (1955), and Dicke (1964).  Wetterich (1995) seems to have been the first to propose that the scalar field in this model for the DE may also fix the DM particle mass. The idea has since been discussed in a considerable variety of contexts (e.g. Damour, Piazza \& Veneziano 2002, references therein, and the references in the following discussion). 

A scalar interaction present only in the dark sector makes the
accelerations of visible and dark matter test particles different,
an effect we will call a ``fifth force.'' The empirical
constraints on this kind of fifth force are considerably weaker
than the constraints from the E\"otv\"os experiment in the visible
sector, as was recognized from the beginning of the modern
discussions, in Damour et al. (1990). 
The first numerical example
we have seen of the effect of this kind of fifth force on the growth of mass density fluctuations in the expanding universe is in Amendola (2000). 
Amendola \&\ Tocchini-Valentini  (2002) 
point out that the fifth force in the
dark sector might have a substantial effect on the relative
distributions of baryonic and dark matter. But we now have
convincing evidence (Bennett et al. 2003 and references therein)
that structure grew out of primeval adiabatic departures from
homogeneity,Ê and good evidence that the growth of the mass density
fluctuations is not very different from the $\Lambda$CDM
prediction, from the consistency within this model between the power spectra of the present distributions of
galaxies and the 3~K thermal cosmic background radiation (the
CBR). Thus one is interested in DM-DE interaction models that can be
adjusted so the fifth force and the evolution of the DE field are
weak enough to fit the now demanding observational constraints,
but strong enough to make an observationally interesting
departure from $\Lambda$CDM. 

The search for models has been influenced by the attractor concept, that the physics may have the property that the astrophysics is insensitive to initial conditions. Peebles \& Ratra (1988) introduced the DE power law potential in equation~(\ref{eq:V}) because it has this attractor property. The physics need not have an attractor, of course, as in the example of Fran\c ca \&\ Rosenfeld (2002), which is based on Wettrich's (1988) potential and suitably chosen initial conditions. We discuss models of this kind in \S 4. 

In the attractor model considered by Anderson \&\ Carroll (1997), and more recently by Comelli, Pietroni \&\ Riotto (2003), the potential of the DE field is the sum of a term linear in the field and  proportional to the DM particle number density (as in equations~[\ref{eq:sf}] and~[\ref{eq:sp}]) and a power law self-interaction term with $\alpha > 0$ in equation~(\ref{eq:V}).  The field is assumed to have been attracted to the minimum of the total potential. Within the models considered in \S 4 this attractor case is unacceptable, because the fifth force is too large. 

Attractor solutions may also be relevant to the dilaton potential. In the scenario considered by Damour \&\ Polyakov (1994) the masses of all particles have minima as a function of the dilaton field at a universal value $\phi _m$; near this minimum the fifth force scales as $(\phi -\phi _m)^2$. They show that comic evolution can cause the present value $\phi$ of the dilaton field to be close to $\phi _m$, thus suppressing the fifth force.

In the class of models we are considering, the contribution to the potential energy of the DE coming from its interaction with the DM has a minimum at zero DM particle mass. This minimum value certainly is not acceptable for the DM in galaxies, but one can imagine that there are two families of DM particles, with
different values of $\phi _\ast$ in equations~(\ref{eq:sf})
and~(\ref{eq:sp}). At large enough DM particle number densities
the DE field would be locked to the zero of the particle mass for the
preponderant family, making these particles relativistic (and
with a mass density that can be acceptably small).
This has the effect of suppressing the fifth force and eliminating
the evolution of the massive DM particle mass and the evolution of the DE density.  This resembles Damour \&\ Polyakov's (1994) ``least coupling,''  but with the difference that in an expanding universe the particle number density must eventually become low enough to release the field. 

We see in this history of ideas a conservative aspect of theoretical physics. The Nordstr\"om action, embodying a variable particle mass, was under discussion before Einstein had completed his general relativity theory of gravity. It reappeared in the 1950s and 1960s, in scalar-tensor generalizations of general relativity expressed in terms of the Einstein action. The scalar-tensor theories were developed to explore Dirac's idea, that the strength of the gravitational interaction may be variable, and these theories served also as a guide to the work of  developing precision tests of gravity physics. The current reappearance of this action is motivated in part by superstring scenarios and in part by the continuing fascination with variable parameters of Nature. Moreover,  the Nordstr\"om action in the form of equation~(\ref{eq:sp}) is equivalent to a Yukawa interaction with a classical scalar field, the form  of which was introduced for very different purposes in meson and weak interaction physics. To be discovered is whether whether this convergence of ideas from particle and gravity physics is a useful guide to observationally significant aspects of the physics of the dark sector. 

\section{Basic Relations}

We begin with the physics of the DM particle model in
equation~(\ref{eq:sp}). Many of the results summarized here have
appeared in one or more of the papers cited above, but we have not seen them all collected or applied. The relation of the
particle model to the field model in equation~(\ref{eq:sf}), which is discussed in \S 3.4, might be considered self-evident, but it should be checked in the present context.

\subsection{Particle and field equations}

We simplify notation in this subsection by setting $\phi _\ast$ to zero (which has the effect of shifting the minimum of $V(\phi )$) and taking $\phi$ to be positive.  Throughout $y$ is positive.  

The particle
action in equation~(\ref{eq:sp}) gives the equation of motion 
\beq
{d\over ds}y\phi g_{\mu \nu}{dx^{\nu} \over ds} = 
{y\phi\over 2} {\partial g_{\rho\sigma }\over\partial x^\mu} 
{dx^{\rho} \over ds} {dx^{\sigma} \over ds} 
+{\partial y\phi\over\partial x^\mu }.
\label{eq:geod} 
\eeq 
We leave the
constant $y$ in this equation because it is useful to note that
the particle four-momentum is $p^\mu = y\phi a\, dx^\mu /ds$. When
spacetime curvature fluctuations can be neglected the equation of
motion is 
\beq 
{d\, a{\bf p}\over dt} = 
{d\over dt}{a y \phi\, {\bf v}\over\sqrt{1 - v^2}}
= - y \sqrt{1 - v^2} {\partial\phi\over\partial {\bf x}},
\label{eq:eqofm} 
\eeq 
where $a(t)$ is the cosmological expansion
factor as a function of the proper world time $t$, 
and the proper peculiar velocity is ${\bf v} = ad{\bf
x}/dt$. When the spatial variation of the DE field $\phi$ may be
neglected the
momentum is conserved;Ê if in addition the proper peculiar
velocity is nonrelativistic, the velocity scales as $v\propto 
1/(a(t)\phi (t))$.

The dark matter that is bound to galaxies and clusters of
galaxies has to be nonrelativistic. These systems are
well described by the weak field limit of 
gravity, where the gravitational potential satisfies \beq \nabla
^2\Phi /a^2 = 4\pi G\rho _b(t)\delta ({\bf x},t).
\label{eq:poisson} \eeq The mean (background) nonrelativistic mass
density is $\rho _b(t)$, and $\delta =\delta\rho /\rho_b$ is the mass
density contrast. We shall write the DE field as 
\beq \phi = \phi _b(t) + \phi _1(\xv ,t), \label{eq:phi1} 
\eeq 
where the mean
background field $\phi _b(t)$ is a function of world time $t$, and
the departure $\phi _1$ from homogeneity may be treated in linear
perturbation theory. (We will also be assuming $| \phi_1| \ll  | \phi_b - \phi_\ast |$, so that there is no risk that the field value passes through zero, which would make the DM transiently relativistic.) In linear theory we may also neglect the term
proportional to ${\bf v}\cdot\nabla\phi$ in equation
(\ref{eq:geod}). With all these 
approximations equation~(\ref{eq:geod}) becomes \beq {d{\bf
v}\over dt} + \left( {\dot a\over a} + {\dot\phi _b\over\phi
_b}\right) {\bf v} = -{1\over a}\nabla\left( \Phi + {\phi
_1\over\phi _b(t)}\right) , \label{eq:eqofmp} \eeq where the dot
is the derivative with respect to time $t$. The
expansion of the universe produces the familiar slowing of the
peculiar velocity in the second term of this equation, while the
evolution of the DE field value produces a term that may increase
or decrease the peculiar velocities. The spatial variation of the
DE field produces the fifth force term in the right hand side of
the equation.Ê This force tends to move DM 
particles so as to minimize their masses $y\phi$.

The DE field equation from the action in equations~(\ref{eq:se})
and~(\ref{eq:sp}) is 
\beq {1\over\sqrt{-g}}{\partial\over\partial
x^\mu } \sqrt{-g}g^{\mu\nu }{\partial\phi\over\partial x^\nu } +
{dV\over d\phi } + {dV_I\over d\phi } =0. \label{eq:feq} 
\eeq 
The term from the interaction with the DM is 
\beqa {dV_I\over
d\phi } & = & y \sum _i {ds_i\over dt}
{\delta ({\bf x} - {\bf x_i})\over\sqrt{-g} } \nonumber\\
& \simeq & y \sum\sqrt{1-v_i^2} \delta ({\bf x} - {\bf x_i})/a^3.
\label{eq:dvdp1} \eeqa
The last expression neglects the effect of
spacetime curvature fluctuations on the DM source term for the DE.

One sees that when the particles are relativistic, $v_i\rightarrow 1$, the source term $dV_I/d\phi$ vanishes. Damour \&\ Polyakov (1994)  express this in terms of an equation of state. We find it convenient to use instead the proper DM particle number density,
\beq 
n({\bf x},t)
=\sum _i\delta ({\bf x} - {\bf x_i})/a^3 =\sum _i\delta ({\bf r} -
{\bf r_i}), \label{eq:numberdensity} 
\eeq 
where $\delta {\bf r}=a\delta {\bf x}$ is a proper relative position.  Thus the source term may be written
\beq 
{dV_I\over d\phi } = y\, n(\xv ,t)\gam .\label{eq:dvdp} 
\eeq 
The brackets signify the mean of the reciprocal Lorentz factor, $\gamma$, for the DM particles. The pressure in this homogeneous gas of DM particles is $p=\gamma y\phi nv^2/3$ and the energy density is $\rho = \gamma y \phi n$, so another form for the source term is $dV_I/d \phi = (\rho - 3p)/\phi$. This is the form derived by Damour \&\ Polyakov (1994).

\subsection{The Fifth Force} 

There is a fifth force in the dark sector, which we analyze for a single nonrelativistic DM family model.ÊThe DM in galaxies and clusters of galaxies is well described by 
the approximations of equations (\ref{eq:poisson})
and~(\ref{eq:eqofmp}), which ignore relativistic corrctions and  the effect of spacetime
curvature fluctuations on the DE field equation. Returning now to the general case when $\phi_\ast$ is non-zero, we can write the spatial mean of the field equation as 
\beq 
{d^2\phi _b\over dt^2} + 3{\dot a\over a}{d\phi _b\over dt} +
\langle dV(\phi )/d\phi\rangle \pm y n_b = 0, \label{eq:phibeq} 
\eeq
where $\phi _b$ is the mean field (eq.~[\ref{eq:phi1}]) and
$n_b(t)$ is the mean particle number density. 
Here and below, the last term has the sign of $\phi - \phi_\ast$.
The departure 
$\phi _1$ from the homogeneous part of the field 
satisfies the equation 
\beq {d^2\phi
_1\over dt^2} + 3{\dot a\over a}{d\phi _1\over dt} - {1\over
a^2}\nabla ^2\phi _1 + {d^2V\over d\phi ^2}\phi _1 \pm y\, n_b\delta
= 0. \label{eq:phi1eqa} 
\eeq 
The DM number density contrast
$\delta n/n$ has been replaced by the mass contrast $\delta
=\delta\rho /\rho$, because, as we argue below, in situations of
interest the fractional perturbation to the field $\phi$ is small
compared to $\delta n/n$.

For the analysis of structure formation at modest redshifts we are
interested in density fluctuations on scales small compared to the
Hubble length, which means the timeÊ derivatives in
equation~(\ref{eq:phi1eqa}) are small compared to the space
derivatives. The term $d^2V/d\phi ^2$ is also small in many cases of interest, so a
useful approximation to the field equation is 
\beq \nabla^2\phi
_1/a^2 = \pm y\, n_b\delta . 
\label{eq:phi1eq} 
\eeq 
It follows from equations~(\ref{eq:poisson}), (\ref{eq:eqofmp})
and~(\ref{eq:phi1eq}) with $\rho _b=y | \phi _b -\phi_\ast| n_b$ that the ratio of
the fifth force to the gravitational force in the dark sector is
\beq 
\beta \equiv {| \nabla\phi _1 |\over | \phi _b - \phi_\ast | \nabla\Phi } = {1\over 4\pi G(\phi _b-\phi _\ast )^2}.
\label{eq:beta} 
\eeq 
The evolution of
the mass density contrast in linear perturbation theory satisfies
\beq {\partial ^2\delta \over\partial t^2} + \left( 2{\dot a\over
a} + {\dot\phi _b\over (\phi _b - \phi_\ast) }\right) {\partial\delta\over\partial
t} = 4\pi G\rho _b(1+\beta )\delta .\label{eq:contrast} 
\eeq 
This differs from the usual expression by the factor $1+\beta$, that
takes account of the fifth force, and the part $\dot\phi _b/(\phi _b - \phi_\ast)
$, from the evolving DM particle mass (as has been discussed by Amendola   \&\ Tocchini-Valentini 2002; Matarrese, Pietroni \&\ Schimd 2003, and others). 

Now we can check that $\delta n/n\simeq \delta\rho /\rho $.
Equation~(\ref{eq:phi1eq}) applied to a particle concentration
with contrast $\delta$ and size $r$ says 
\beq  {\delta m\over m} = {\phi _1 \over ( \phi
_b - \phi _\ast )} \sim {- y n_b r^2 \delta \over | \phi _b - \phi_\ast | }  ,
\eeq   where $H$ is the Hubble parameter.  The sign of the mass shift $\delta m/ m$ is opposite to the sign of $\delta$, consistent with the attractive nature of the fifth force.  Using the condition that the DM mass density is not greater than the total, we find that the fractional mass shift is  
\beq  { | \delta m| \over m} \lesssim \beta (Hr)^2 | \delta |. \label{massshift} 
\eeq
This is small because $\beta$ cannot be much larger
than unity, and we are interested in density fluctuations on
scales small compared to the Hubble length.

\subsection{A Relativistic Dark Matter Family}

The DM bound to galaxies and clusters of galaxies has
to be nonrelativistic, with $\gam$ close to unity, so the source
term~(\ref{eq:dvdp}) due to the scalar field interaction with such DM is simply proportional to the particle number density. But there may be more than one DM family, and $\phi$ may be drawn close to the zero of the source term belonging to one of the families, causing that family to be relativistic. Here we consider the behavior of this piece of $dV_I/d\phi$ as a function of the field value, under the simplifying assumption that the scalar field is spatially homogeneous. In \S 5, on a model with two DM families, we consider the response of $\phi$ to the source term and the effects of inhomogeneities.  

The velocity of a DM particle in this relativistic family is
\beq
v = {k/a\over\sqrt{y^2\phi ^2 + k^2/a^2}},
\label{eq:velocity}
\eeq
where the comoving wavenumber of a DM particle is $k$ and the proper peculiar momentum is $k/a$. The initial distribution of comoving wavenumbers is determined by the DM particle production process.  We will discuss the evolution of $\phi$ in \S 5. For the purpose of this subsection we are concerned with the case 
$y |\phi | \lesssim k/a\ll H^{-1}$. The first inequality means the DM particles are relativistic; the second means the discussion of  the evolution of the mass distribution can ignore time derivatives of $\phi$ and the expansion of the universe.  In this approximation the energy of a DM particle is conserved as it propagates through space and encounters DE field gradients. Therefore the local value of the DM particle velocity changes in response to changes in the local value of the DM particle mass, as determined by the local value of the DE field. With this in mind, we can rewrite equation~(\ref{eq:dvdp}) as
\beq
{dV_I\over d\phi } = y\, n(t) \, \langle y\phi \, (y^2\phi ^2 +
k^2/a^2)^{-1/2}\rangle _k,
 \label{eq:vi} 
\eeq 
where the average is over the distribution of particle wavenumbers in the relativistic family.  When $\phi$ passes through zero (or more generally, through
the zero of the particle mass) in a close to homogeneous way it
does not greatly affect the distribution of momenta $k/a$, but it
makes the DM peculiar motions relativistic. That causes the source term~(\ref{eq:vi}) to vary smoothly from $dV_I/d\phi =y\, n$ at 
$y \phi \gg k_{\rm eff} /a$ to $dV_I/d\phi = -y\, n$ at 
$y \phi \ll -k_{\rm eff}/a$, where the  effective comoving wavenumber $k_{\rm eff}$ is defined by the average in equation~(\ref{eq:vi}). 

As a final remark, we note that, according to the action principle, the equation of motion is derived by extremizing the action~(\ref{eq:sb}) with the particle orbits held fixed.  Thus although $V_I$ may be written as a function of $\phi$, $n$, and the momenta $k/a$, the source term $dV_I/d\phi$ is not the same as the derivative of $V_I$  with respect to $\phi$ at fixed $n$ and $k/a$.  In the derivation of the field equations from the Lagrangian~(\ref{eq:sf}), which will be discussed in \S 3.4, the independent field $\psi$ is held fixed as $\phi$ is varied, leading to equation~(\ref{eq:dvdp}).

\subsection{The field action}

Here we consider, in the limit where the particle de Broglie wavelengths are small compared to the length scale of variation of $\phi$, the equivalence of the particle action in equation~(\ref{eq:sp}) to the spin-1/2 fermion field action in equation~(\ref{eq:sf}) and the boson field action in equation~(\ref{eq:sb}), for the purpose of deriving particle orbits and the source term for $\phi$.  For brevity we again take $\phi$ to be positive and
suppress $\phi _\ast$. 

If the parameters $y$ and $\phi$ in the Yukawa interaction term in
equation~(\ref{eq:sf}) are positive the particles created by the field
$\psi$ manifestly have positive mass $y\phi$. A chiral rotation by
$\pi$ changesÊ the sign of $\bar\psi\psi$ without changing the
kinetic part of the field equation, so when $y\phi$ is negative
the chiral rotation yields the usual sign for theÊ mass in the
Dirac equation. Alternatively, we can leave the ``wrong'' sign for
the mass when $\phi$ is negative, and note by the chiral
transformation argument that the solutions to the field equation
in this case make $\bar\psi\psi$ negative, meaning
$y\phi\bar\psi\psi$ is never negative. This condition leads to the
prescription for the absolute field value in the particle
action~(\ref{eq:sp}).

In the field action (eq.~[\ref{eq:sf}]) the source term for 
the DE field is Ä
\beq 
dV_I/d\phi = y \bar\psi\psi = y\, n\sqrt{1-v^2} , \label{eq:dvdpf}
\label{eq:source}
\eeq 
when $y$ and $\phi$ are positive. 
The last expression follows because $\bar\psi\psi$ and
$n\sqrt{1-v^2}$ both are scalars, and we know they are the same --
up to the sign -- in the nonrelativistic limit.\footnote{One can
check these arguments by writing down plane wave solutions to
the Dirac equation. The DE source term in equation~(\ref{eq:source}) can be derived from the
free quantum field operator for $\psi$, apart from the standard problem with the zero-point contribution to the particle number operator.} This 
agrees with the source term in equation~(\ref{eq:dvdp}) in the
particle model.

The second step in demonstrating the equivalence of the
actions~(\ref{eq:sf}) and~(\ref{eq:sp}) is to check the equation
of motion of wave packets. An easy way to proceed uses the
commutator of the particle momentum operator with the Dirac
Hamiltonian $H=\hat \alpha\cdot {\bf p} +\hat \beta y\phi$, \beq [{\bf p},H]
= - i y \hat \beta\nabla \phi . \eeq It
follows that the time derivative of the expectation value of the
momentum is \beq {d{\bf p}\over dt} = -y\nabla \phi \int d^3r \psi
^\dagger \hat \beta\psi . \label{eq:eqcheck} \eeq But the last factor is
the integral over $\bar\psi\psi$, which we know is the reciprocal
of the Lorentz factor for a single particle wave packet, as in
equation~(\ref{eq:dvdpf}). Equation~(\ref{eq:eqcheck}) thus agrees with
the rate of change of momentum in the particle model in
equation~(\ref{eq:eqofm}).

For completeness let us check the relation between the
momentum and velocity of the wave packet. An easy way 
is to use a WKB approximation. In considering the
motion of the wave packet we can ignore the time evolution of
$\phi$. The interesting part of the spatial variation in one
dimension of a wave function with energy $\epsilon$ is 
\beq
\psi\sim\exp i\left(\int ^x\sqrt{\epsilon ^2 - y^2\phi ^2} dx -
\epsilon t\right) , 
\eeq which means the momentum defined by the gradient
operator is 
\beq
p = \sqrt{\epsilon ^2 - y^2\phi ^2},
\label{eq:momentum}
\eeq
as usual. The
velocity of a wave packet constructed as a linear combination of
these energy eigenstates follows from the stationary point of the
exponential: $v=p/\epsilon$, again as usual. And these two results
give the standard relation between momentum and velocity of a
particle of mass $y\phi$. On can also check that, when the time evolution of $\phi$ and the expansion parameter $a(t)$ can be neglected, the time derivative of equation~(\ref{eq:momentum}) agrees with the rate of change of momentum in equation~(\ref{eq:eqofm}).

Similar arguments show that in the limit of short de Broglie wavelengths the boson DM model in equation~(\ref{eq:sb}) also reduces to the point particle model (apart from the problem of the zero point contributions to $\phi ^2$ and $\chi ^2$), with DM  particle mass $y | \phi - \phi_\ast |$.  

\section{Models with one dark matter family}

In the models presented in this section, the DE potential is the power law
form in equation~(\ref{eq:V}), with positive or negative power law
index $\alpha$, and there is one family of nonrelativistic DM
particles with $\phi _\ast =0$, meaning the DM particle mass is
$y$ times the DE field value. For the purpose of this preliminary
exploration we neglect the mass in baryons, so the matter density
parameter is \beq \Omega _mH_o^2 = {8\over 3}\pi G\rho _b(t_o) =
{8\over 3}\pi Gy\, \phi _b(t_o)n_b(t_o). \label{eq:omegam} \eeq
Here and below the subscript $o$ means the present value. The
Hubble parameter at the present world time $t_o$ is $H_o$.
Throughout we use $\Omega _m=0.3$ and we assume space curvature
vanishes. \footnote{For simplicity in this exploratory discussion of observational constraints we do not adjust the value of $\Omega_m$ to take account of the fifth force.  In a theory with a fifth force $\beta$ enters different measures of $\Omega_m$ in different ways.  For instance, the relative motions of dark matter halos depend on the product $\Omega_m (1+\beta)$,  whereas weak lensing depends on the fifth force only indirectly, through whatever effect the scalar field has on the angular size distance.  The dynamics of ordinary matter are not directly affected by the fifth force in the dark sector. }

\begin{deluxetable}{rcrrcc}
\tablecaption{Numerical results for one DM family. \label{table1}}
\tablewidth{0pt}
\tablehead{
\colhead{$\alpha$} & 
\colhead{$G^{1/2}\phi _{i}$} & 
\colhead{$\kappa\qquad $} &
\colhead{$ {\phi _{\rm eq}\over |\phi _o|}$} &
\colhead{${\delta _o\over\delta _{\Lambda{\rm CDM}}}$} &
\colhead{${l_{\rm peak}\over l_{\Lambda{\rm CDM}}}$}
}
\startdata
Ê --2 &Ê 0.95 & 1.5E+01 &ÊÊÊÊ --Ê &ÊÊ --Ê &ÊÊ --Ê \\
Ê --2 &Ê 1.00 & 1.4E+00 &ÊÊÊÊ --Ê &ÊÊ --Ê &ÊÊ --Ê \\
Ê --2 &Ê 2.00 & 3.0E--02 &ÊÊ 1.22 &Ê 1.18 &Ê 1.18 \\
Ê --2 &Ê 4.00 & 5.7E--03 &ÊÊ 1.04 &Ê 1.04 &Ê 1.04 \\
ÊÊÊ 4 &Ê 0.50 & 4.7E--03 &ÊÊÊ --Ê &ÊÊÊ -- &ÊÊ --Ê \\
ÊÊÊ 4 &Ê 1.00 & 3.5E--03 &ÊÊ 2.17 &Ê 2.88 &Ê 2.02 \\
ÊÊÊ 4 &Ê 2.00 & 6.8E--01 &ÊÊ 1.19 &Ê 1.18 &Ê 1.17 \\
ÊÊÊ 4 &Ê 4.00 & 1.8E+01 &ÊÊÊ 1.04 &Ê 1.04 &Ê 1.04 \\
ÊÊÊ 6 &Ê 0.50 & 1.0E--02 &ÊÊÊ -- &ÊÊ --Ê &ÊÊ --Ê \\
ÊÊÊ 6 &Ê 1.00 & 1.3E--02 &ÊÊ 1.33 &Ê 1.37 &Ê 1.29 \\
ÊÊÊ 6 &Ê 2.00 & 2.0E+00 &ÊÊÊ 1.18 &Ê 1.17 &Ê 1.16 \\
ÊÊÊ 6 &Ê 4.00 & 2.8E+02 &ÊÊÊ 1.04 &Ê 1.03 &Ê 1.04 \\
\enddata
\end{deluxetable}

We use the dimensionless variables \beq \tau = H_ot,\qquad f =
G^{1/2}\phi _b , \label{eq:tau} \eeq in terms of which
equation~(\ref{eq:phibeq}) for the mean field $\phi _b$ is \beq
{d^2f\over d\tau ^2} + {3\over a}{da\over d\tau }{df\over d\tau }=
{\alpha\kappa\over f^{\alpha + 1}} - {3\over 8\pi }\left( a_o\over
a\right) ^3{\Omega _m\over f_o}, \label{eq:f} \eeq and the
Friedman equation for the expansion rate is \beq \left( {1\over
a}{da\over d\tau }\right) ^2 = \Omega _m {f\over f_o}\left(
a_o\over a\right) ^3 + {8\pi\over 3}\left[ {1\over 2}\left(
df\over dt\right) ^2 + {\kappa\over f^\alpha }\right]
,\label{eq:a} \eeq where the dimensionless parameter representing
the constant $K$ in the power law potential $V(\phi )$ is \beq
\kappa = KG^{1+\alpha /2}/H_o^2. \label{eq:kappa} \eeq The present
values $f_o$ and $a_o$ of the field and expansion parameter appear
in the combination $f_o/a_o^3$, which is the unknown final
condition for given initial conditions.

At high redshift the first term on the right hand side of
equation~(\ref{eq:f}), which represents $dV/d\phi$, is relatively
small. When this term may be neglected the first integral of
equation~(\ref{eq:f}) is, apart from the decaying term, 
\beq
{df\over d\tau } = - {3\Omega _m\tau\over 8\pi f_o} \left(
a_o\over a\right) ^3. \label{eq:dfdt} 
\eeq 
At $z>z_{\rm eq}$,
where the expansion is dominated by radiation, the expansion
factor varies $a\propto\tau ^{1/2}$, and equation~(\ref{eq:dfdt})
says the departure from the initial value of $f$ grows as $\tau
^{1/2}$. At lower redshift where the expansion is
matter-dominated, $a\propto\tau ^{2/3}$, the departure grows as
$\log\tau$ in the approximation of equation~(\ref{eq:dfdt}). At
still lower redshifts $dV/d\phi$ may be important, and we need a
numerical solution.

We commence the numerical solution at a fixed initial time
corresponding to equality of mass densities in matter and
radiation in the $\Lambda$CDM model. We start with an arbitrary
choice for the initial field value $f_{i}=G^{1/2}\phi _{i}$. 
The initial value of $df/d\tau$ is taken from
equation~(\ref{eq:dfdt}). The final field value (and hence energy
density) has to be consistent with $da/d\tau =a$ at the present
epoch. We achieve this 
by iteratively adjusting $\kappa$. (We found this more convenient
than choosing $\kappa$ and seeking the initial field value.)
Having solved numerically for $f(\tau )$, we can find the epoch
$z_{\rm eq}$ at equal mass densities in matter and radiation.

Table~1 lists parameters and present values of some quantities of
interest for solutions with three choices for the value of
$\alpha$, omitting numbers that are so far off the $\Lambda$CDM
model prediction as to seem uninteresting. The second column is
the initial field value, expressed in units of the Planck mass.
The third column is the value of $\kappa$ required for a
consistent solution. The fourth column is the ratio of the field
value at $z_{\rm eq}$ to the present value.Ê Because we are
assuming $\phi_* = 0$, this is the ratio of DM particle masses
then and now. The redshift 
at equality scales in proportion to this ratio.

Figures 1 and 2 show the evolution of the DE field in solutions
with $\alpha = -2$ and $\alpha = 6$. The latter look much like the
solutions for $\alpha =4$ entered in Table 1. Since we have set
$\phi _\ast=0$, the DE field in solutions with $\alpha <0$ is
drawn toward zero. At $\alpha = -2$ and the smallest initial field
value -- listed in the first line of the table and shown as the
dotted curve in Fig. 1 -- the field has passed through zero
slightly before the present epoch. Among other undesirable
consequences, this would have made the DM transiently
relativistic, driving the DM out of the halos of galaxies. The
slightly larger initial field value in the second line of the
table, with the appropriate adjustment of $\kappa$, removes this
problem: in this case, shown as the dashed curve in Figure 1, the
DM particle mass has not yet passed through zero, but that will
soon happen and the halos will be disrupted.

When $\alpha$ is positive the potential for $\phi$ has a minimum
away from zero. If the initial value of $\phi$ is small enough the
field relaxes to this minimum by the present epoch. This is seen
in our solution with the smallest initial field value, listed in
the fourth entry from the bottom of Table~1 and plotted as the
dotted curve in Fig. 2. In this case the field oscillates about
and approaches the minimum of the potential.Ê This solution is 
unacceptable, however, because the
relatively small field values produce a large fifth force on the
dark matter, which substantially 
enhances the growth of mass density fluctuations, as we
discuss next.  At the two largest initial field values in Table 1,
the solutions for $\alpha =6$ are well away from the minimum of
the potential. They look much like the solutions for $\alpha = 
-2$, and as we show next, produce only a modest effect on the
evolution of mass density fluctuations.\footnote{Anderson \&\ Carroll 
(1997) and  Comelli, Pietroni \&\ Riotto (2003)
consider the case that  the scalar field sits at the minimum of
the  effective potential. They adopt the same particle coupling and potential (with $\alpha  >0$) as in the examples in this section.
However when the scalar field is at the minimum of the effective potential -- as in an attractor scenario -- there is an unacceptably rapid evolution of the DM particle mass subsequent to 
decoupling. This evolution, and the fifth force, can be suppressed  by choosing a large value of $\phi_{\ast} \sim m_{Pl}$
in equation~(\ref{eq:beta}). Our acceptable-looking cases are 
not attractor solutions: they are sensitive to the initial value
of $\phi$, and their success relies in part  on the assumption that $\phi$ is far
from its value at the minimum of the potential.}

Figures~3 and~4 show numerical solutions to
equation~(\ref{eq:contrast}) for the evolution of the mass density
contrast $\delta (t)$ in linear perturbation theory. The solutions
are normalized to a common initial value at redshift $z=1300$,
roughly the epoch of decoupling, and they are multiplied byÊ the
redshift factor $1+z=a_o/a(t)$ to scale out the main trend of the
evolution. The solution for the $\Lambda$CDM model, with the same
value of $\Omega _m$, is plotted as the short dashed curves in
both figures. The solution for $\alpha =-2$ and $f_{i}=G^{1/2}\phi
_i=0.95$ is not plotted because equation~(\ref{eq:contrast}) does
not take account of the transient relativistic motions of the DM
particles. The fifth column of Table~1 lists the ratio of the
growth factor since decoupling, $\delta _o/\delta _{\rm dec}$, to
the prediction of the $\Lambda$CDM model.

At the two largest initial field values and all three choices of
$\alpha$, the growth of density fluctuations is close to the
$\Lambda$CDM prediction. In the solution for $f_{i}=1$ and $\alpha
=-2$, plotted as the long dashed curves in Figs. 1 and~3, the
growth of the density contrast is about four times that of the
$\Lambda$CDM model. That is ruled out by the consistency of the
CBR temperature anisotropy and the large-scale fluctuations in the galaxy distribution
within the $\Lambda$CDM model. For
$f_{i}=1$ and $\alpha =6$ the density fluctuation growth factor is
more than a factor of two different from $\Lambda$CDM at redshift
$z=10$, but happens to be fairly close at the present epoch (as
one sees by comparing the long and short dashed curves in Fig. 4).
This solution is challenged by the position of the peak of the CBR
fluctuation spectrum, however, as we now discuss.

\begin{figure}
\plotone{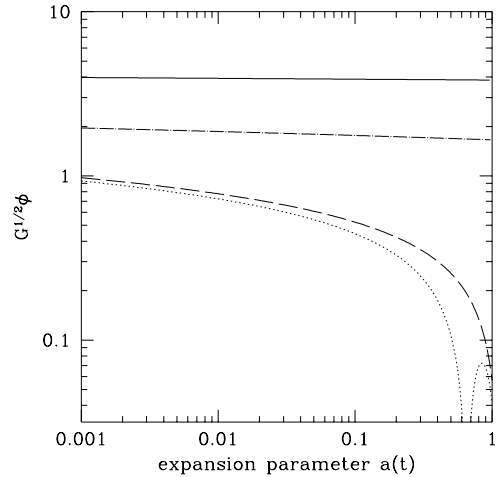}
\caption{Evolution of the DE field in models with the power law
exponent $\alpha = -2$ in the potential (eq.~[\ref{eq:V}]). The
solutions are fixed by the initial value of $f=G^{1/2}\phi$ listed
in the first four rows of Table 1. The initial values are close to
the field values at the left-hand edge of the plot.
\label{fig1}}
\end{figure}

\begin{figure}
\plotone{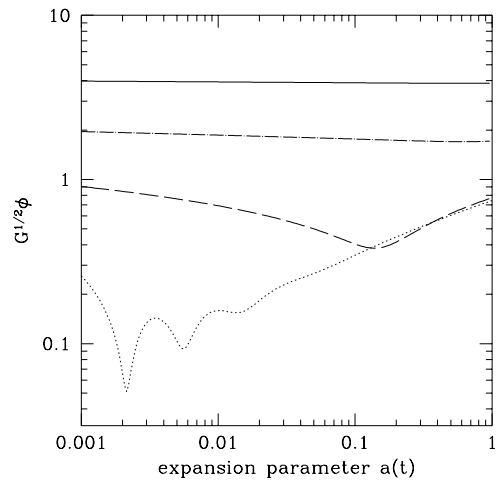} 
\caption{The
same as Fig. 1 for solutions with $\alpha =6$. The initial values
of $\phi$ are listed in the last four entries in Table 1.
\label{fig:fig2} }
\end{figure}

\begin{figure}
\plotone{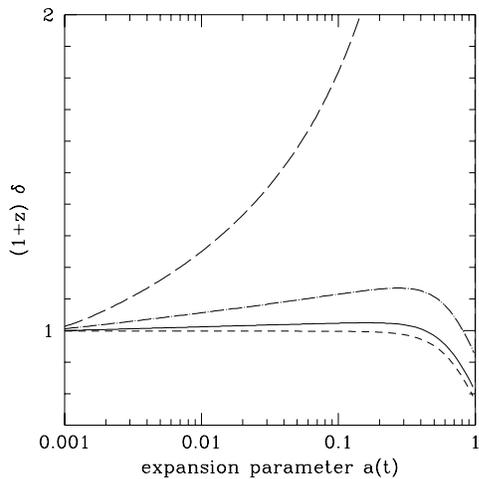}
\caption{Evolution of the mass density contrast in linear
perturbation theory in models with $\alpha = -2$. The density
contrast has been multiplied by the redshift factor $1+z$ to scale
out the evolution when the expansion is matter-dominated. The
short dashed curve is the solution for the $\Lambda$CDM model. The
line types of the other curves match Fig. 1, where the initial
field values are close to what is plotted at the left side of the
figure. \label{fig:3} }
\end{figure}

\begin{figure}
\plotone{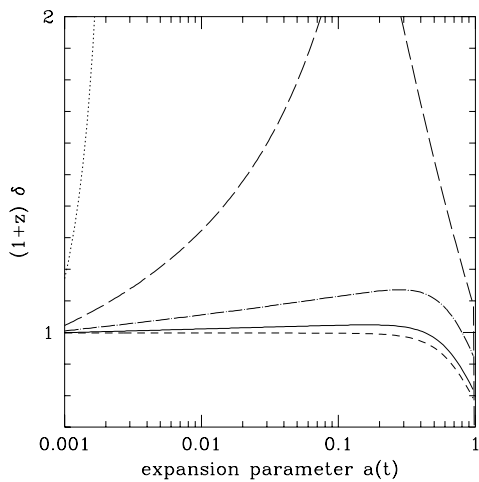} 
\caption{The
same as Fig. 3 for solutions with $\alpha = 6$ \label{fig:4} }
\end{figure}

We estimate the angular scale of the peak of the CBR temperature
anisotropy power spectrum as follows. Since the physical
wavelength of the mode that produces the peak of the fluctuation
spectrum is set by the Hubble length at $z_{\rm eq}$, the
wavenumber at the peak varies with the model parameters as \beq
{k_{\rm peak}\over a_o}\sim {a_{\rm eq}\over a_o}{1\over t_{\rm
eq}} \propto {a_o\over a_{\rm eq}}\propto {f_{\rm eq}\over f_o}.
\eeq The second step follows because the expansion time is
inversely proportional to the square of the temperature at $z_{\rm
eq}$, that is, $t_{\rm eq}\propto a_{\rm eq}^2$, and the last step
follows because the redshift at equal mass densities in radiation
and DM varies as $f_{\rm eq}/f_o$ through the evolution of the DM
particle mass. The peak of the angular power spectrum of the CBR
temperature is at spherical harmonic number $l_{\rm peak}\sim
k_{\rm peak}r$, where the angular size distance $r=\int dt/a$ is
integrated from decoupling to the present epoch. Thus the ratio of
the spherical harmonic index $l_{\rm peak}$ in the model to the
predicted index at the peak in the $\Lambda$CDM model with the
same cosmological parameters is \beq {l_{\rm peak}\over
l_{\Lambda{\rm CDM}}}\simeq {f_{\rm eq}\over f_o}{r\over
r_{\Lambda{\rm CDM}}}. \label{eq:l} \eeq This ratio is listed in
the last column of Table~1.

The model with $\alpha =6$ and $G^{1/2}\phi
_{i}=1$, whose present density fluctuations happen to be close to
the $\Lambda$CDM prediction, puts the peak of the CBR temperature
fluctuation spectrum at angular scale $\sim 30$\%\ smaller than
$\Lambda$CDM, which likely is unacceptable. At $G^{1/2}\phi
_{i}=2$ the peak is shifted from $\Lambda$CDM by about 16\% ,
which may be tolerable within the uncertainties allowed by the
other cosmological parameters that determine the value of $l_{\rm
peak}$. A closer analysis of the joint distribution of allowed
values of $\phi _i$ and the cosmological parameters seems
inappropriate in this preliminary exploration.Ê The point we wish
to demonstrate is that there is a range of initial field values
that produce a significant but acceptable departure from the
behavior of the $\Lambda$CDM model.

\begin{figure}
\plotone{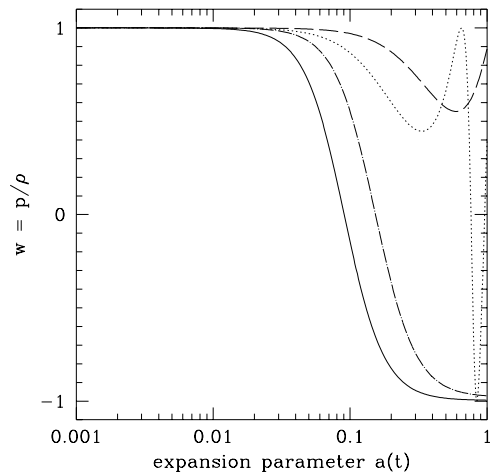}
\caption{The
evolution of the equation of state parameter (Fig.~[\ref{eq:w}])
in solutions with $\alpha = -2$. The line types match Fig. 1.
\label{fig:5} }
\end{figure}

\begin{figure}
\plotone{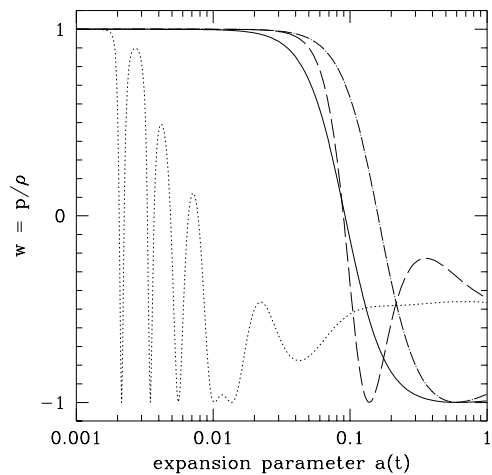}
\caption{The
evolution of the equation of state parameter in solutions with
$\alpha = 6$. The line types match Fig. 2. \label{fig:6} }
\end{figure}

The evolution of the DE density nowadays is characterized by an
effective DE equation of state. We define the effective
pressure $p_{{\rm eff}}$ by the expression for local energy
conservation, 
\beq 
{d\over dt}(\rho _{\rm DM} + \rho _{\rm DE}) = -3{\dot a\over
a}(\rho _{\rm DM} + \rho _{\rm DE} + p_{{\rm eff}}). 
\eeq 
The ratio of the effective pressure to the DE density isÊ 
\beq 
w = {p_{{\rm eff}} 
\over\rho _{\rm DE}} = - \, {V -\dot\phi ^2/2Ê 
Ê+( \nabla \phi )^{2}/2 \overÊ V + \dot\phi ^2/2 
Ê+( \nabla \phi )^{2}/2Ê }. \label{eq:w} \eeq 
Figs. 5 and~6 show the evolution of the equation of state
parameter $w$ in our numerical solutions, which neglect the
gradient energy density in equation~(\ref{eq:w}). The complicated
behavior of $w$ in the solutions with the two smallest initial
field values is of no interest because the models are not viable.
At the two larger initial field values the parameter is close to
constant at $w\simeq -1$ in the range of redshifts reached by the
SNeIa observations. At higher redshifts $w\simeq +1$, because in
these solutions the 
DE energy is dominated byÊ $\dot\phi ^2/2$, but at high redshift the DE
density is well below the DM mass density.

\section{Cosmologies with two dark matter families}

Here our field Lagrangian is 
\beqa \label{eq:sf2} 
L & = &Ê \phi _{,\nu }\phi
^{,\nu } /2
- V(\phi +\phi _s) \nonumber \\
& & +i\bar\psi\gamma\cdot\partial\psi
+ i\bar\psi _s\gamma\cdot\partial\psi _s \nonumber \\
Ê& & - y (\phi _\ast - \phi  )\bar\psi\psi
- y_s\phi\bar\psi _s\psi _s, 
\eeqa
where for definiteness we choose $y$,
$y_s$, $\phi _s$ and $\phi _\ast$ to be positive constants.  The mean number densities are $\bar n$ in the first family, with wave functions
$\psi$, and $\bar n_s$ in the second family. If $\phi$ is never
close to zero or to $\phi _\ast$ then there are two
nonrelativistic DM families, a situation that can be equivalent
to what we discussed in the previous section. A new possibility
offered by this model is that 
at high particle number density the DE field is ``locked'' to $\phi \simeq 0$ to minimize the effective potential of the family with the larger value of the Yukawa coupling constant times the number density, 
making that family
relativistic. If $y_s n_s>y n$ in equation~(\ref{eq:sf2}), and $V(\phi)$ is subdominant, the particle
mass in the massive DM family is fixed to $m=y\phi _\ast$, and the
DE density is fixed to $\rho _{\rm DE} = V(\phi _s)$, so the DE
behaves like Einstein's cosmological constant. We show in \S 5.1
that the locking can also substantially suppress the fifth force.
In \S 5.2 we comment on the complicated behavior when the number
density in the second family becomes small enough to allow the DE
field to evolve.

\subsection{A relativistic family in the dark sector}

We make several approximations that simplify the analysis of the
fifth force when the DE field is close to the minimum of the
potential of one family.

First, we assume the length scale of the density fluctuations of
interest is much smaller than the Hubble length, so the dark
matter particles can relax to near time-independent equilibrium.
Second, we assume that at high redshift, when the DM number densities are large, the scalar field relaxes to the minimum of the particle potential.  We take $y_sn_s\gg y\bar n$, so the mean value of the scalar field is close to zero. This means there is a nonrelativistic DM family with mean mass density $y\phi _\ast\bar n$ together with a relativistic second family. Third, it
is reasonable to take it that the space distribution of the
relativistic family is close to homogeneous. We can see how this
comes about by considering the distribution of positions and
momenta $p=\sqrt{\epsilon ^2 - y^2\phi (\xv )^2}$ in single
particle phase space. If the distribution has relaxed to become
nearly independent of time and a function only of the energy
$\epsilon$, then the space number density distribution for the second family 
particles with energy $\epsilon$ is \beq n_s(\xv ) \propto\int
d^3p\,\delta (\epsilon - \sqrt{\phi ^2+p^2}) \propto p/\epsilon =
v. \label{eq:microcanonical} \eeq This must be averaged over the
distribution of energies. But we see the key and familiar point,
that when the second family is relativistic, that is, $v$ is close
to unity, the space distribution is close to homogeneous. 

To get a viable model
we have to choose parameters so the energy density in the
relativistic family is small enough to avoid spoiling light
element production. The typical energy $\epsilon _{\rm eff}$ of a
relativistic second family particle is dominated by its momentum,
which scales with the expansion of the universe as $a(t)^{-1}$,
because the expansion stretches the de Broglie wavelengths. Thus,
as long as the second family remains relativistic, its mean energy
density is $\bar\rho _s = 
n_s\epsilon _{\rm eff}\propto a(t)^{-4}$, as usual for relativistic particles. The ratio of
the mean energy densities in 
the two DM families is 
\beq
 {\bar\rho _s\over\bar\rho } \sim
{\epsilon _{\rm eff}n_s\over y\phi _\ast\bar n} 
 ={y_sn_s\over y\bar n}{\epsilon _{\rm eff}\over y_s\phi _\ast}.
\label{eq:ratio} 
\eeq 
The first factor in the last expression must be larger than unity, say by a factor of 100, to keep $\phi$ locked to the second family even as density concentrations in the first family develop. To avoid affecting the standard model for the origin of the light elements we want the energy density in the relativistic second family to be small compared to the thermal background radiation, meaning the present density is $ (\bar\rho _s/\bar\rho )_0\la 10^{-4}$. Thus initial conditions  for the particle momenta must be such that $\epsilon _{\rm eff}\sim k_{\rm eff}/a_0\lesssim 10^{-6}y_s \phi _\ast$.

Under the above assumptions,  and neglecting $ dV/d\phi$ for the moment, the DE
field equation when  $\phi$ is near zero is
\beq 
\nabla ^2\phi /a^2 = y_s^2n_s|\phi |/\epsilon _{\rm eff} -
Êy\bar n(1+\delta ) .
\label{eq:aa} \eeq
In the first source term we
have written the relativistic correction (eq.~[\ref{eq:dvdp}]) in
terms of the effective mean particle energy, $\epsilon
_{\rm eff}$, as in
equation~(\ref{eq:vi}):
\beq
 \gam \equiv y_s |\phi (\xv )| /\epsilon
_{\rm eff}. \label{eq:eff} 
\eeq
We have dropped the time
derivatives of $\phi$ because 
we are assuming the field is locked to a value near zero. The
number density $n_s$ in the second (relativistic) family is nearly 
independent of position. The density contrast in the nonrelativistic DE is $\delta (\xv ,t)$. On scales small compared to the
Hubble length the second family particles see a nearly static
potential, so the particle energy $\epsilon
_{\rm eff}$ is conserved and thus independent of position along its trajectory.

The space average of equation~(\ref{eq:aa}), which neglects $dV/d\phi$, gives the mean field
value, 
\beq \phi _b = \epsilon _{\rm eff} y\bar n/(y_s^2n_s).
 \eeq
This relation in equation~(\ref{eq:eff}) reproduces the condition that the mean inverse Lorentz factor for the second family
satisfies 
\beq \gam = y\bar n/(y_sn_s ) \ll 1. \eeq

The departure from the mean of equation~(\ref{eq:aa}) (with $\phi _b>0$ since we are taking $dV/d\phi =0$ for the moment) is
\beq
\nabla ^2\phi _1/a^2 = y_s^2n_s\phi _1/\epsilon _{\rm eff} - y\bar
n\delta . 
\label{eq:ab} \eeq 
The Fourier transform is \beq \phi
_1(\kv ) = {y\bar n\delta (\kv ) \over k^2/a^2 + y_s^2n_s/\epsilon
_{\rm eff} }.
\eeq
The Green's function thus has a Yukawa
form, $\propto r^{-1}\exp{ -r/r_5}$, with cutoff length $r_5 =
\sqrt{\epsilon _{\rm eff}/y_s^2n_s}$.  This scales with time as $r_5\propto a(t)$, so the comoving cutoff length is constant. 

If the Hubble parameter $H$
is dominated by the mass density $y \bar n \phi_{*}$ in the
nonrelativistic DM, the cutoff 
length satisfies 
\beq (Hr_5)^2 \simÊ{ G\epsilon _{\rm eff}\phi _\ast y\bar n
\over y_s^2n_s } \sim {1\over \beta}
{\bar\rho _s\over \bar\rho }
\left( y \bar n\over y_s n_s\right) ^2. 
\eeq
As argued above, the last two factors are at most $10^{-4}$ and $(10^{-2})^2$, so that 
\beq (Hr_5)_0 \la 10^{-4} \beta^{-1/2} .
\label{eq:Hr5}
\eeq 
To produce $r_5(a_0)$ of order the Hubble radius would require $ \beta \sim 10^{-8}$ and thus $\phi_\ast \sim 10^4 m_{Pl}$, which is disagreeably large on theoretical grounds.  On the other hand, $\beta =1 $ implies $r_5(a_0) \la  1 $ Mpc which is well within the non-linear  clustering length, and so requires closer analysis to test.

\subsection{Late Time Transition}

The behavior of $\phi$ when the number densities become small
enough to allow the field to evolve depends on the self-interaction potential $V(\phi )$. In linear perturbation theory for the field, the condition that the second family particle number density is large enough to hold $\phi$ constant and close to zero is 
\beq dV/d\phiÊ = \pm y_sn_s\gam +  y\bar n (1+\delta ).
\label{eq:xx}
\eeq 
This assumes $dV/d\phi >0$ at $\phi\simeq 0$.  The negative sign in the first term on the right hand side 
applies when $\phi >0$, and the positive
sign when the particle number densities are small enough to allow 
$dV/d\phi $ to pull $\phi$ to a slightly negative value. 
Because the second family is relativistic, $n_s$ is nearly
homogeneous, as we have discussed. This means the 
reciprocal Lorentz factor (eq.~[\ref{eq:eff}]) must be a function of
position, balancing the irregular DM mass distribution in the
first family represented by the number density contrast 
$\delta (\xv ,t)$.  We will check below that even when the mass density contrast is non-linear,  equation~(\ref{eq:xx}) is a good approximation for the perturbation to the field.

We
first consider the case where $dV/d\phi$ can be neglected, so
equation~(\ref{eq:xx}) is
\beq
y_sn_s\gam = y \bar n(1+\delta ). \label{eq:aaa} 
\eeq 
There comes a time when the value $\delta _{\rm max}$
of the density contrast within the strongest concentrations of the
massive DM family is large enough to satisfy $y_sn_s = y\bar
n(1+\delta _{\rm max})$. This 
forces the second family to become nonrelativistic in the
neighborhood of $\delta _{\rm max}$. Further expansion of the universe
increases the density contrasts, causing $\phi$ in the vicinity of a DM mass
concentration to increase. When $y_s\phi >\epsilon _{\rm eff}$, second family
particles are pushed out of the regions of first family
concentrations because their energy is not sufficient to allow them to have such a large mass. If this rearrangement is happening on length scales much
smaller than the Hubble length, the DE field equation is dominated
by the spatial derivatives, and we have 
\beq \nabla ^2\phi /a^2 =
y_sn_s(\xv, t) \gam - y \bar n (1+\delta ), \label{eq:bb} 
\eeq
for the case that $dV/d\phi$ can be neglected.
As indicated, 
the second family number density $n_s$ is now a function of position,
because less energetic particles are excluded from the concentrations of the first
family. The spatial mean of equation~(\ref{eq:bb}) says 
that the inverse Lorentz factor averaged over all second
family particles satisfies 
\beq \langle \gam 
\rangle = y\bar n/(y_s\bar n_s ).
\label{eq:cc} \eeq
Since we are assuming the ratio on the right hand side is smaller
than unity, the pools of relativistic second family
particles in the regions between the concentrations of
nonrelativistic DM are always able to hold theÊ field value close
to zero in the voids between the concentrations of galaxies. 

We have been assuming the field value within a concentration of the
first family is less than $\phi _\ast$, so these particles are
nonrelativistic. To estimate the extent of the mass shift of first family particles, consider a concentration of
$N\sim\bar nR^3$ nonrelativistic DM particles drawn from an initially homogeneous patch of size $R$ into a concentration with density contrast $\delta$ over a size
$r$. The typical shift in the DE field value averaged over this concentration is
$\phi _r \sim yN/r$.   Assuming the expansion rate is mainly due to the
mass density in this family so the Hubble parameter satisfies 
$H^2\sim G y \bar n \phi _\ast$, we get 
\beq 
{\phi _r\over\phi _\ast } \sim
{(HR)^2\over G\phi _\ast ^2} {R\over r} \sim
\beta (Hr)^2 \delta. \label{eq:dd} 
\eeq 
The fractional shift in the scalar field value is largest in the largest mass concentrations.  For example, the density contrast in a large galaxy is about $\delta = 10^6$ at $r=10$ kpc, which gives $\phi _r/\phi _\ast \approx 10^{-5} \beta $, and in a rich cluster at $r=2$ Mpc where $\delta \simeq 100$,  $\phi _r/\phi _\ast \approx  10^{-4} \beta $.  That is, as long as $\beta$ is not large the first family particle masses are only
slightly perturbed by the spatial variation of $\phi$, and the force on massive DM particles from the gradient of $\phi$ is well approximated as $\beta$ times the gravitational attraction (eq.~[\ref{eq:beta}]). 

Finally, let us briefly consider what happens when $dV/d\phi$ in
equation~(\ref{eq:xx}) is positive and large enough to pull $\phi$ to negative values. This makes the second family first 
become nonrelativistic in the voids, where the first family
number density is low. When this happens the field in the voids
moves toward the minimum of $V$, at $\phi = -\phi _s$. If 
$y_{s}\phi _s\gg\epsilon _{\rm eff}$ then the DE field pushes the
second family out of the voids.
Since the mean particle number densities are decreasing as the universe expands, the
potential $V$ must eventually pull the DE field away from zero
everywhere, producing a second family of nonrelativistic dark matter. 

When the DE field is no longer locked to the zero of mass for the second family there is a fifth force, but it need not parallel the peculiar gravitational attraction of the dark matter because the space distributions of the two families may differ. 

A model in which the lock on the DE field has broken before the present epoch and produced a large fifth force within concentrations of galaxies would not be acceptable, but a model with a large fifth force in the voids between the concentrations of large galaxies might be quite interesting,  as a way to understand why the voids are so empty.

\section{Concluding Remarks}

We have explored physical processes in models
where the DM particles have a Yukawa coupling to a scalar field that can be the source of the DE.  Our central conclusion is that parameters and initial conditions 
in such models can be chosen so that the model is
viable but significantly different from the standard $\Lambda$CDM 
cosmology. It might be useful to follow this up by considering whether the interesting range of initial field values, just somewhat larger than the Planck mass for the models in \S 4,  could naturally follow from models for the very early universe.  It would also be useful to know whether the number density of DM particles required to give the observed value of $\Omega_{m}$ can be
naturally understood, for example by gravitational production.Ê 

As illustrated in 
\S 4 for a single DM family, the constraints on cosmological
parameters derived by fitting the model for the dark sector to the observations 
can differ from what is obtained from fitting to the standard
model.  Such alternative models are 
therefore useful as foils to $\Lambda$CDM,  for the purpose
of evaluating the empirical constraints on the cosmological
parameters

We have also considered two-family models, that can lead to a
rich and interesting cosmology.  As discussed in \S 5, initial
conditions can be 
chosen so the DE field is locked to the zero of mass for
the more numerous family, and remains so up to the present epoch. 
This removes the evolution of the DE field and the evolution of the mass of the nonrelativistic DM particles, and it can 
suppress the fifth force in the dark sector. This is an example
of how a dynamical model for the DE can be observationally
indistinguishable from Einstein's cosmological constant. The
situation changes when the DE particle number density becomes low enough to free the DE field. When this
happens the behavior can become quite different from the standard
model.

Models of the type studied here, in which 
one or possibly several families of dark matter 
have masses set by their interaction with a 
dynamical scalar field, are a useful cautionary 
example of another point: the empirical evidence on how the universe 
has been evolving up to now may be a rather deceptive
guide to its physics or to its future. 

\acknowledgments

We benefitted from discussions with Thibault Damour, Bharat Ratra, and  Alex Vilenkin. PJEP acknowledges support from the NSF.Ê GRF thanks Princeton University's Departments of Astronomy and of PhysicsÊ for their hospitality; her research is also supported in part by NSF-PHY-0101738.

\end{document}